\documentclass[usenatbib,useAMS,a4paper,fleqn]{mnras}
\usepackage{amssymb}
\usepackage{newtxtext,newtxmath}
\usepackage[T1]{fontenc}
\usepackage{ae,aecompl}
\usepackage{times}
\usepackage{microtype}
\usepackage{verbatim}
\usepackage{graphicx}
\date{Accepted 2020 August 17. Received 2020 July 2; in original form 2019 November 27}
\pubyear{2019}

\usepackage[multidot]{grffile}

\begin{document}

\label{firstpage}

\pagerange{\pageref{firstpage}--\pageref{lastpage}}

\title[Making massive stars in the Galactic Centre]
{Making massive stars in the Galactic Centre via accretion onto low-mass stars within an accretion disc}
\author[Davies \& Lin]{Melvyn B. Davies$^{1}$ \& Doug N. C.  Lin$^{2}$  \\
$^{1}$Lund Observatory,  Department of Astronomy and Theoretical Physics,  Box 43, SE--221 00, Lund, Sweden \\
$^{2}$UCO/Lick Observatory, Board of Studies in Astronomy and Astrophysics, University of California, Santa Cruz, CA 95064, USA}

\date{Accepted 2020 August 17. Received 2020 July 2; in original form 2019 November 27}

\maketitle

\begin{abstract}
The origin of the population of very massive stars observed within $\sim 0.4$ pc
of the supermassive black hole in the Galactic Centre is a mystery.
Tidal forces from the black hole would likely inhibit {\it in situ} star formation
whilst the youth of the massive stars would seem to exclude formation elsewhere followed by transportation
(somehow) into the Galactic centre. Here we consider a third way to produce
these massive stars from the lower-mass stars contained in the nuclear stellar
cluster which surrounds the supermassive black hole. 
A passing gas cloud can be tidally shredded by the supermassive black hole
forming an accretion disc around the black hole. 
Stars embedded within this accretion disc will accrete gas from the disc
via Bondi-Hoyle accretion, where the accretion rate onto a star, $\dot{M}_\star
\propto M_\star^2$. This super-exponential growth of accretion can lead to 
a steep increase in stellar masses, reaching the required 40-50 M$_\odot$ in some cases.
The mass growth rate depends sensitively on the stellar orbital eccentricities 
and their inclinations. The evolution of the orbital inclinations and/or their
eccentricities as stars are trapped by the disc, and their orbits 
are circularised, will increase the number of massive stars produced.
Thus accretion onto low-mass stars can lead to a top heavy stellar mass function
in the Galactic Centre and other galactic nuclei. The massive stars produced 
will pollute the environment via supernova explosions and potentially produce 
compact binaries whose mergers may be detectable by the LIGO-VIRGO 
gravitational waves observatories.
\end{abstract}

\begin{keywords}
Galactic centre; Stars
\end{keywords}

\section{Introduction}

Observational studies of the very centre of our own Milky Way Galaxy reveal a very 
unusual stellar population \citep[see e.g.][]{2010RvMP...82.3121G}.
A disc of very massive stars ($10-60$ M$_\odot$)
are found within 0.4 pc of the supermassive black hole 
 \citep[e.g.][]{2000MNRAS.317..348G,2003ApJ...590L..33L}.
About 200 stars above 20 M$_\odot$ are seen  \citep[cf. Fig. 3 of ][]{2010ApJ...708..834B}.
The high masses of these stars require them to be young (around 6 Myr). 
They seemingly have a top-heavy IMF, with $dN/dm \propto m^{-0.45}$, i.e.\ much
flatter than a normal IMF. In addition, stellar dynamics at the galactic centre
limits the number of low-mass stars formed together with the massive
stars \citep[][]{2007ApJ...654..907A}.

Forming these stars from gas {\it in situ} is problematic owing to their
proximity to the supermassive black hole (and its large tidal forces which would shred
a normal GMC at sub-pc distances). 
An infalling cluster, if formed relatively locally, could have been tidally shredded
producing the population of young, massive stars \citep{2001ApJ...546L..39G}.
However, such a process would likely leave the stars 
further away from the central supermassive black hole than observed.

In this paper we consider a third option. A GMC passes close to the 
supermassive black hole and is shredded forming an accretion disc.
The observed Fermi bubble \citep[][]{2010ApJ...724.1044S} could well have formed due to a recent 
accretion episode from a disc formed in such a way.
It has been suggested that star formation within such a disc, through fragmentation
due to self-gravity within the disc, could produce 
the observed young, massive stars 
\citep[e.g.][]{2003ApJ...590L..33L,2005A&A...437..437N,2006MNRAS.372..143N,2007MNRAS.374..515L}.
Here we consider a variation on this model, where we make use of the 
pre-existing  lower-mass stars within the nuclear stellar cluster.
These stars may become more massive as they 
accrete material from the disc \citep{2003ApJ...590L..33L}.
As we will see in this paper, given suitable disc properties, and favourable
stellar orbits, the accretion rate may be sufficient to grow Solar-like stars
into very massive stars comparable to those observed.

 This paper is set out as follows.
We describe the  accretion disc model used in Section 2.  In Section 3, we
derive the time evolution of a star's mass as it accretes gas from the accretion disc
and present timescales for  growth due to the accretion disc presented
in Section 2. In Section 4 we discuss results, noting the possible
importance of migration within the disc, as well as considering how 
the accretion of gas might be inhibited once the star reaches very 
large masses (where the sum of stellar and accretion luminosities 
may exceed the Eddington luminosity). We present our conclusions in Section 5. 

\section{Accretion disc model}

We follow here the approach given by \citet{2020ApJ...889...94M}.
Namely we consider an accretion disc with material accreting at a constant
rate onto the supermassive black hole, $\dot{M}_{\rm smbh}$. Both the Toomre
$Q$ parameter, which relates to whether the disc would be vulnerable to forming
self-gravitating lumps, and the viscosity parameter $\alpha$ are assumed to be
constant throughout the disc.

The Eddington mass accretion rate $\dot{M}_{\rm Edd} = L_{\rm Edd} / \eta c^2$,
where the Eddington luminosity is given by $L_{\rm Edd} = 4 \pi G M_{\rm smbh} m_{\rm p} c / \sigma_{\rm T}$,
$\sigma_{\rm T}$   being the Thompson cross section for electron scattering and $m_{\rm p}$ being the
proton mass. Here we take $\eta = 0.1$, and thus obtain

\begin{equation}
\dot{M}_{\rm smbh} = \lambda \dot{M}_{\rm Edd} = 0.088 \lambda \left( {M_{\rm smbh} \over 
4 \times 10^6 {\rm M}_{\odot} } \right) {\rm M}_\odot \ {\rm yr}^{-1}
\end{equation}
The surface density of the disc $\Sigma$ is  given by the expression

\begin{equation}
\Sigma =  { \dot{M}_{\rm smbh} \over 2 \pi r v_{\rm r} }
\end{equation}
where the radial inflow in the disc $v_{\rm r} = \alpha h^2 (GM_{\rm smbh} /r)^{1/2}$
and $h$ is the ratio of disc scale height $H$ to radius (i.e.\,$h = H/r$), which is given by 
\begin{equation}
h^3 \simeq { Q \over 2 \alpha } { \dot{M}_{\rm smbh}  \over M_{\rm smbh} \Omega }
\end{equation}
where the disc angular frequency $\Omega = ( G M_{\rm smbh} / r^3 )^{1/2}$.
The surface density of the disc can be rewritten in terms of the critical surface density
required for the disc to be self-gravitating

\begin{equation}
\Sigma = { \Sigma_{\rm c} \over Q}= {h \over Q }{ M_{\rm smbh} \over \pi r^2} = {1 \over (2 \alpha)^{1/3} Q^{2/3} }
{ M_{\rm smbh}^{2/3} \dot{M}_{\rm smbh}^{1/3} \over \pi r^2 \Omega^{1/3} }
\end{equation}
Thus our accretion disc is therefore characterised by the constants, $\lambda$, $\alpha$ and $Q$,
with $h \propto r^{1/2}$ and $\Sigma \propto r^{-3/2}$.

We take $M_{\rm smbh} = 4 \times 10^6$ M$_\odot$ for the mass of the supermassive black hole.
We consider here an accretion rate at 10 per cent Eddington (i.e.\,$\lambda = 0.1$), with the disc on 
the edge of the self-gravitating instability ($Q=1.0$), the instability driving the viscosity parameter
$\alpha$ to  also be of order unity  \citep{1995ARA&A..33..505P}.
The evolution of such discs has been considered in detail \citep[e.g. ][]{2001ApJ...553..174G,2007MNRAS.379...21N}.
They may fragment into low-mass lumps. Such lumps may coagulate to ultimately
form more massive stars  \citep[e.g. ][]{2007MNRAS.374..515L}. Alternatively, energy released
via accretion onto the lumps may heat the disc, increasing the disc scale height , and thus $h$, and
therefore increasing the value of $Q$ such that the disc is no longer unstable due
to its self-gravity \citep[e.g. ][]{2006MNRAS.372..143N}. Given that our disc will contain 
pre-existing stars from the nuclear stellar cluster which happen to find themselves in the 
disc plane, the accretion energy released as gas flows onto these stars will also heat the
disc. Radiation from stars above and below the disc will also increase the disc temperature.
One can therefore imagine a disc teetering on the edge of being unstable. We 
take our disc model here as a simple limiting case. In reality the disc may be slightly thicker,
and thus have a slightly lower density and a slightly higher sound speed, which in turn lead
to slightly lower accretion rates. However, as we will see shortly, our disc model can produce
the massive stars seen on timescales somewhat less than 10 Myr in some cases, thus massive
stars are still likely to be produced within 10 Myr even if the accretion rates are somewhat reduced.

We consider a disc of radius $R_{\rm disc} = 3 \times 10^{18}$ cm, 
and thus obtain a total disc mass of $M_{\rm disc} = 3.16 \times 10^5$ M$_\odot$.
From equation (1), we see that our disc has an accretion rate of $0.0089 {\rm M}_\odot \  {\rm yr}^{-1}$,
thus the disc lifetime $\tau_{\rm disc} \simeq 36$ Myr. It is also very thin, 
with $h \sim  10^{-3} - 10^{-2}$.  Though as discussed above, heating via accretion onto stars 
and from the radiation of stars above and below the disc, is likely to increase the thickness
of the disc (and thus $h$).

The disc we use here is broadly consistent with the observed Fermi bubble having been formed during an accretion
episode onto the supermassive black hole from such a disc 
\citep[e.g.][]{2010ApJ...724.1044S,2011MNRAS.415L..21Z}. Such a disc could be produced
by the infall and tidal shredding of a GMC. 
The infall may be the result of a collision between two gas clouds, resulting in a merged
object possessing relatively little angular momentum. 
We note that the resulting infalling 
cloud may already in some cases be fragmenting into denser lumps
as a result of the collision
\citep[e.g.][]{2008Sci...321.1060B,2009MNRAS.394..191H,2013MNRAS.433..353L}.

\section{Growth by accretion}

We can place stars on orbits within our accretion disc and follow their
evolution as they accrete gas via Bondi-Hoyle accretion.
Here we will make a number of simplifying assumptions:  we do not consider
the time evolution of the accretion disc, or the eccentricity and inclination of
a star's orbit.

For a star of mass $M_\star$ passing through a gas of density $\rho$ with a relative
speed $v$, the Bondi-Hoyle accretion rate is given by  \citep[e.g.][]{2004NewAR..48..843E}

\begin{equation}
\dot{M}_\star = { 4 \pi G^2  \rho \over \left( v^2 + c_s^2 \right)^{3/2} } M_\star^2
\end{equation}
where the density can be approximated using $\rho \simeq  \Sigma  / 2 H$
and $c_s$ is the sound speed of the gas, which is given through the relation
$h \sim (c_s / v_\phi) $ where $v_\phi = \sqrt{G M_{\rm smbh}/R}$.
One can also make the approximation, $v \simeq e v_\phi$ where 
$e$ is the eccentricity of the star's orbit. Integration of an entire orbit
reveals that $v$ varies between $\sim 0.5 - 1.0  \ e v_\phi$ with a time-average
probably close to about $\sim 0.75 \  e v_\phi$.
Solving $\dot{M}_\star = K M_\star^2$, assuming K is a constant, we
obtain:

\begin{equation}
M_{\rm \star, b} = {M_{\rm \star, a}  \over 1 - K M_{\rm \star, a} \left( t_{\rm b} - t_{\rm a} \right)}
\end{equation}
where $M_{\rm \star, a}$ and $M_{\rm \star, b}$ are the masses of the star at times  
$t_{\rm b}$ and $t_{\rm a}$ (the initial time) respectively.
As can be seen from above, the stellar mass increases to very large values
once $K M_{\rm \star, a} ( t_{\rm b} - t_{\rm a}) \sim 1$. This leads us
to introduce the {\it interesting timescale}, $t_{\rm int}$ which is given by:

\begin{figure}
\vspace{-1.0 cm}
\begin{center}
\includegraphics[width=3.0in]{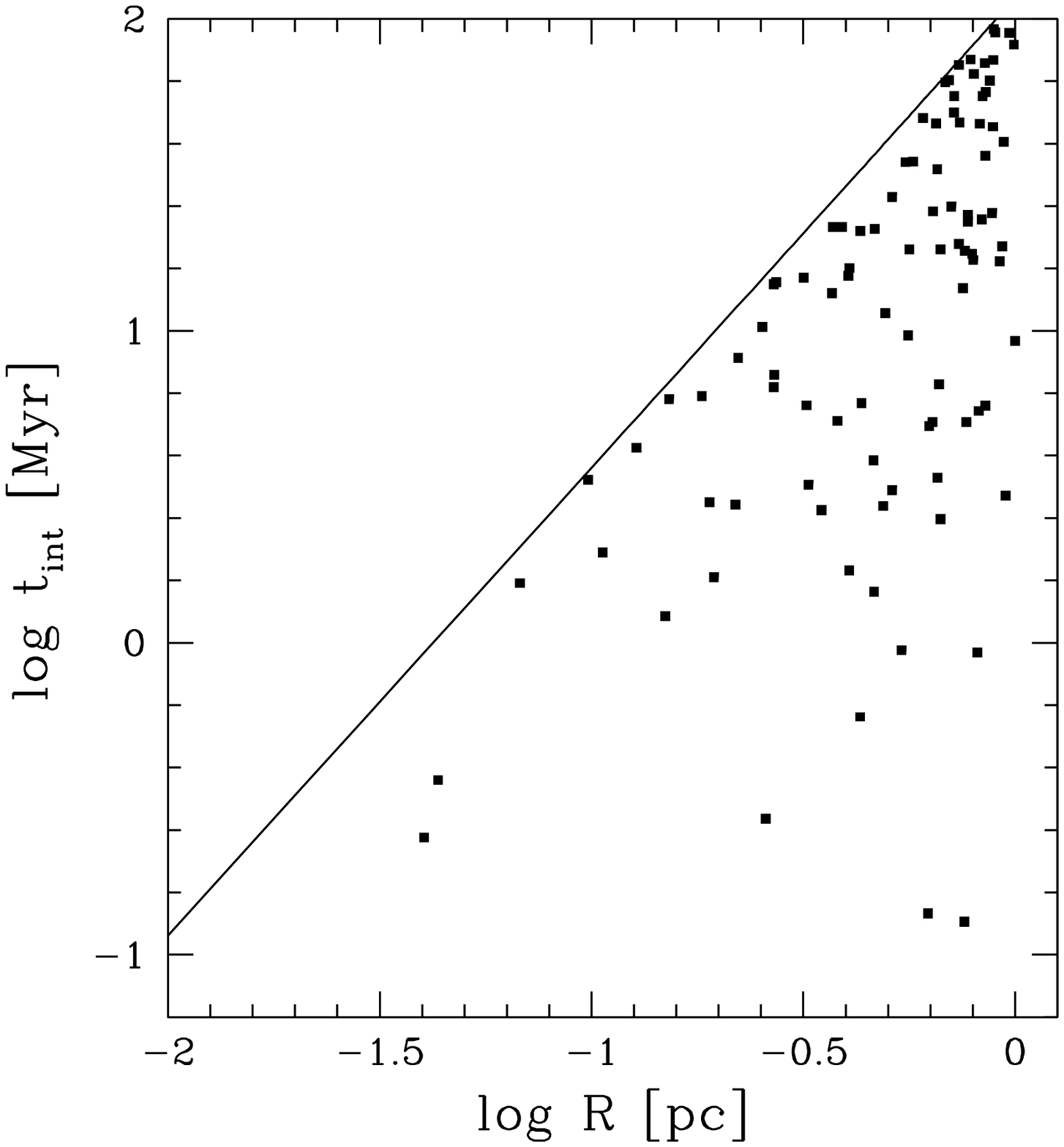} 
 \vspace*{-2.0 cm}
 \caption{The Log of the timescale for growth in stellar mass, $t_{\rm int}$ plotted as a function 
 of the Log of the radius, $R$. The line is for stars having an eccentricity $e=0.2$. The dots are for
 a Monte Carlo population having an eccentricity drawn from a thermal distribution but
 with $e < 0.2$. All stars, and the line drawn, are placed within the accretion disc.}
   \label{mbdavies_figure1}
\end{center}
\end{figure}

\begin{figure}
 \vspace{-1.0 cm}
\begin{center}
\includegraphics[width=3.0in]{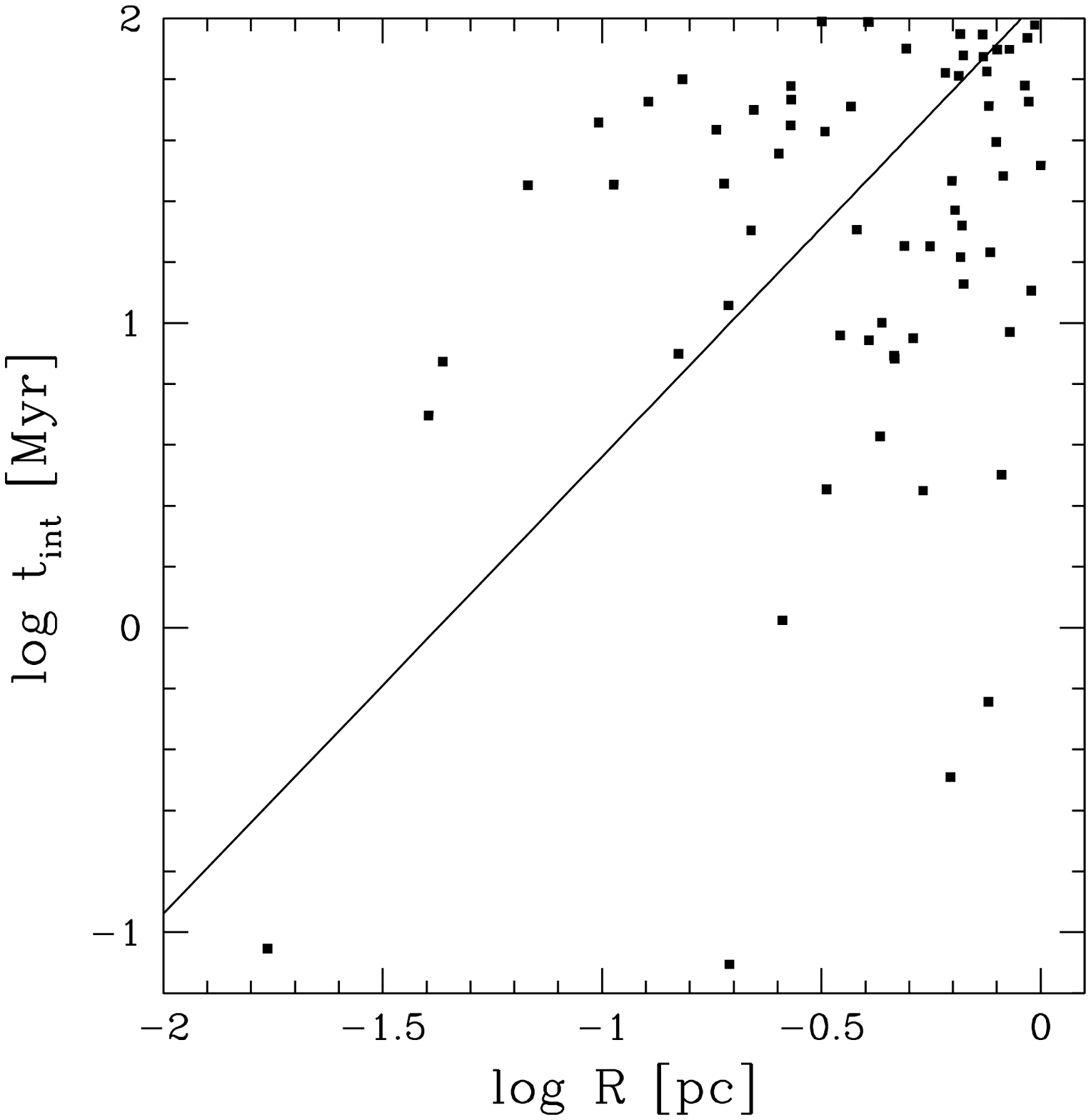} 
 \vspace*{-2.0 cm}
 \caption{As for Fig.\ 1 but with the inclinations of the Monte Carlo population of stellar
 orbits drawn randomly for inclinations, $i < 0.1$ radians}
   \label{mbdavies_figure2}
\end{center}
\end{figure}

\begin{equation}
t_{\rm int} = {1 \over K M_{\rm \star, a} }
\end{equation}
where $M_{\rm \star, a}$ is the initial stellar mass. It worth pointing out a simple
extrapolation from the initial accretion rate would substantially underestimate the 
total increase in mass. For example, for $M_{\rm \star, a} = 1 $M$_\odot$,
and an initial accretion rate, $\dot{M}_{\star, a} = 0.1$ M$_\odot$ Myr$^{-1}$, one would
formally reach an infinite mass in 10 Myr, 
whereas by a simple extrapolation, one would have estimated 
that the mass gained in 10 Myr would be only
  $\Delta M_\star \simeq 0.1 \times 10 = 1$ M$_\odot$.

We initially only consider stars lying within the accretion disc. 
We assume the stars are radially distributed with a number density
$n = k r^{-\gamma}$ where here we take $\gamma=1.75$. 
In Fig.\ 1 we show $t_{\rm int}$ as a function of radius for orbits
with eccentricity $e = 0.2$. One can see from this figure, 
that growth timescales can be interestingly short
(we need to grow in a few Myr or less as this is
the evolutionary timescale for the massive stars we wish to produce).
We also produce a population of 100 
stars, each located within the disc, but having eccentricities drawn
from a thermal distribution (i.e.\  $df= 2e de$)  between $e=0$ and 
$e=0.2$. One can see from this plot that $t_{\rm int}$ is a sensitive function
of eccentricity. As our disc is very thin (and therefore relatively cold),
the denominator in Equation (5) is dominated by thespeed
of the star relative to the gas, hence at a given radius,  $\dot{M}_\star \propto e^{-3}$.

In Fig.\ 2 we consider the case where not all stars are within the accretion
disc. Explicitly we randomly sample inclinations up to 0.1 radians. 
Most of these stars will only spend a fraction of their orbit within the disc.
Neglecting for now the effect of stellar capture by the disc, we can
estimate the increase in $t_{\rm int}$, as
$t_{\rm int, inc} = t_{\rm int} \times i \times R/H$, where $i$ is the inclination
of the star's orbit. In reality
we expect a number of these stars to settle into the disc
thus shortening their growth timescale. We plot $t_{\rm int}$
as a function of their orbital eccentricity in Fig.\ 3, clearly showing
the dependence on eccentricity.

We evolve the masses of our 100 stars shown in 
Fig.\ 2 for 10 Myr. We do not let
any mass exceed $40$ M$_\odot$. In reality the massive
stars will have a spread in masses due to their detailed 
individual evolutions but here we will simply use this maximum mass
as a label for the production of very massive stars. 
We find that about 25 per cent of our stars become
very massive. In Fig.\ 4 we
plot the final stellar mass $M_{\star, b}$ as a function of orbital
eccentricity. We see here again the critical dependence of the growth
on the orbital eccentricity. All stars with $e < 0.1$ become very massive
whereas the growth is negligible for $e > 0.15$. In Fig.\ 5 
we plot $M_{\star, b}$ as a function of radius. We see here how,
interestingly, the formation rate of massive stars is relatively independent
of their  position within the accretion disc.

\begin{figure}
\vspace*{-1.0 cm}
\begin{center}
\includegraphics[width=3.0in]{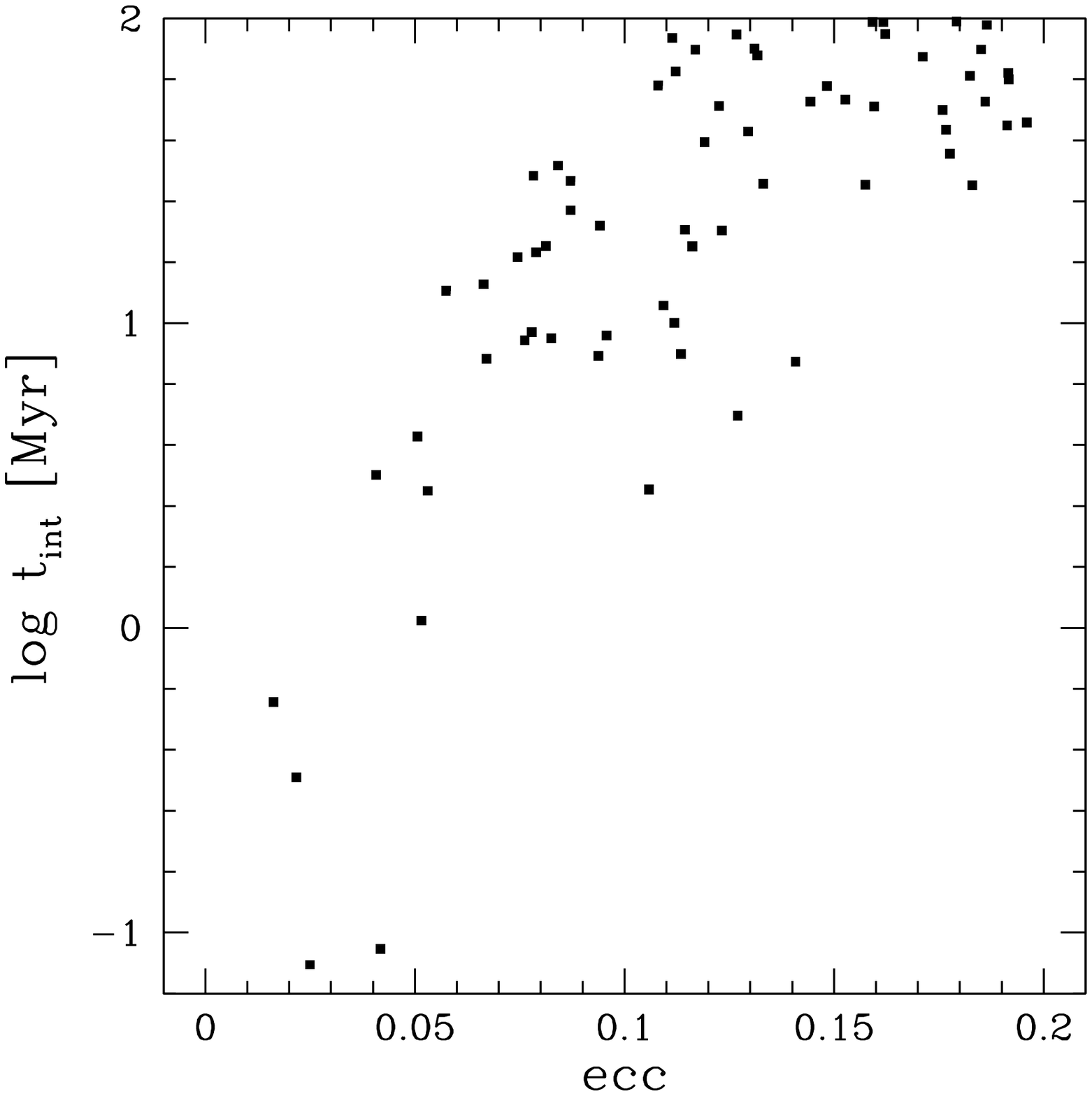} 
 \vspace*{-2.0 cm}
  \caption{The Log of the timescale for growth in stellar mass $t_{\rm int}$ plotted as a function 
 of eccentricity of the stellar orbits, for the systems shown in Fig. \ 2.}
   \label{mbdavies_figure3}
\end{center}
\end{figure}

\begin{figure}
\vspace{-1.0 cm}
\begin{center}
\includegraphics[width=3.0in]{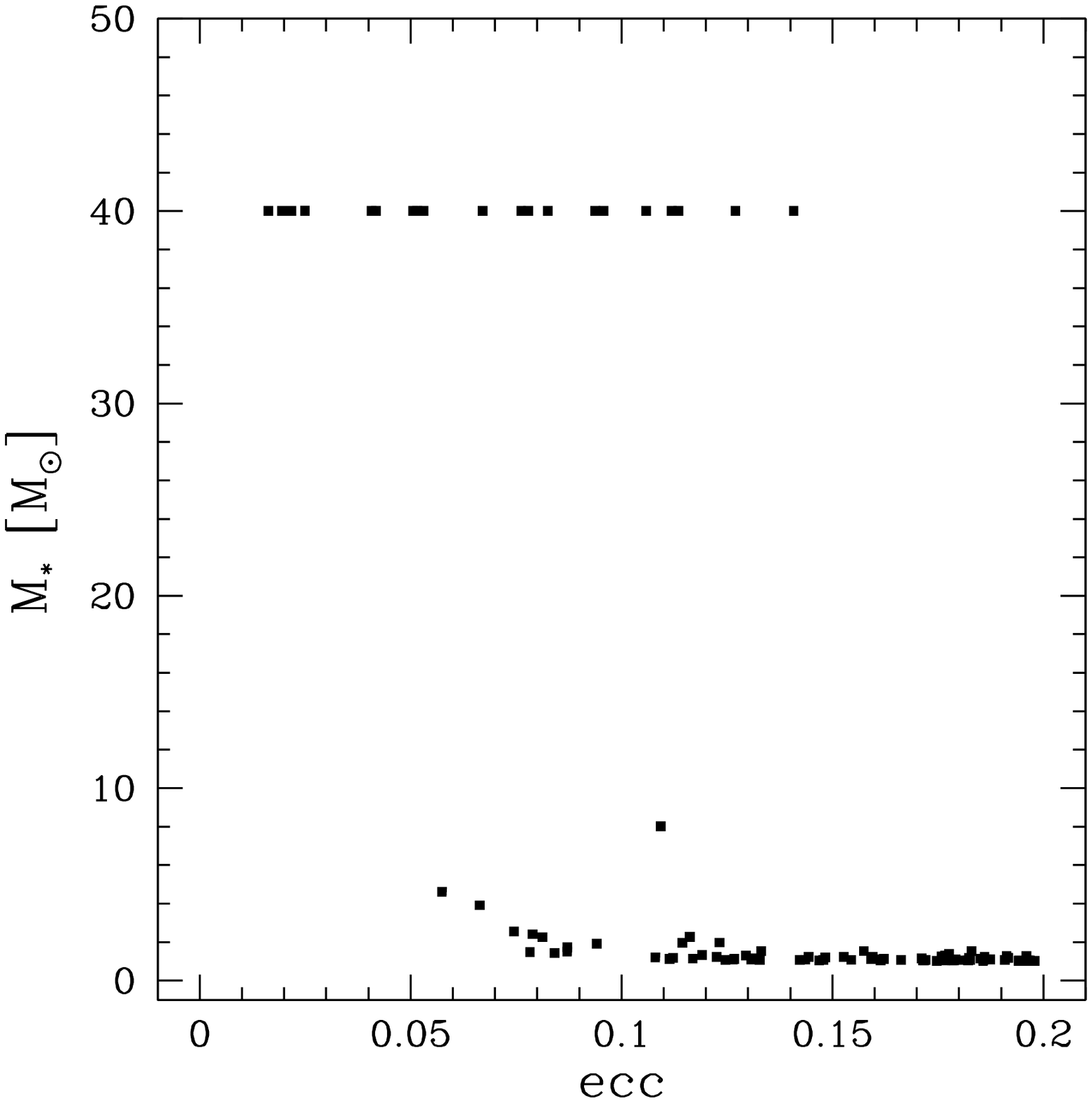} 
 \vspace{-2.0 cm}
 \caption{Final stellar masses, $M_\star$ after 10 Myr of evolution (as described in 
 Section 3) plotted as a function of orbit eccentricity.}
   \label{mbdavies_figure4}
\end{center}
\end{figure}

\begin{figure}
\vspace*{-1.0 cm}
\begin{center}
\includegraphics[width=3.0in]{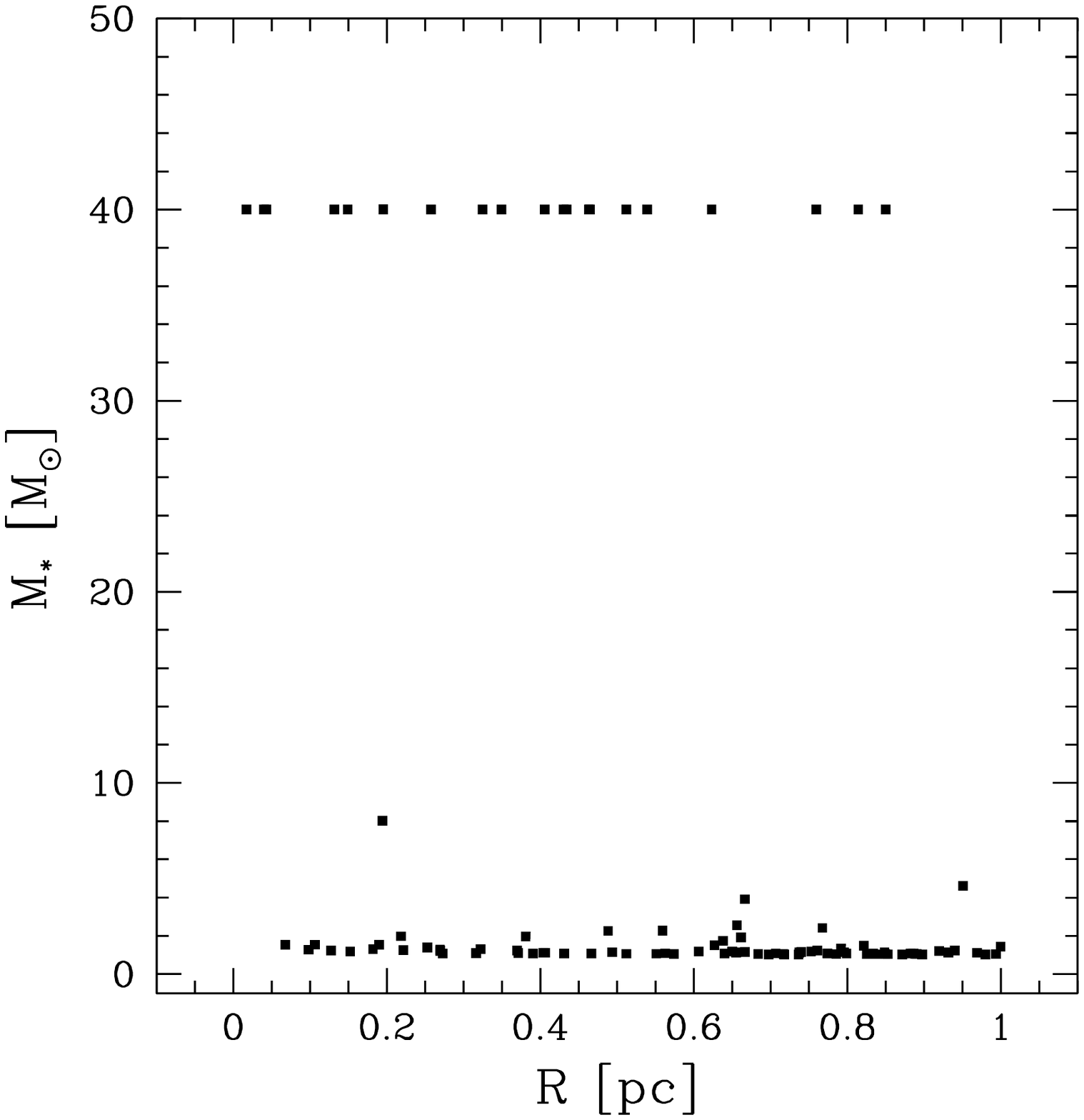} 
 \vspace*{-2.0 cm}
  \caption{Final stellar masses, $M_\star$ after 10 Myr of evolution (as described in 
 Section 3) plotted as a function of  radius.}
    \label{mbdavies_figure5}
\end{center}
\end{figure}

\section{Discussion}

\medskip

As we saw in Section 1, the observed number of very massive stars is around 200. Our calculation
presented here with a population of 100 stars, $0 < e < 0.2$,  inclination $i < 0.1$, 
produced around 20 very massive stars. How many massive stars would be produced
within a disc from a reasonable nuclear stellar cluster population?
If we assume isotropic orbits, then the fraction of stars with inclination $i < 0.1$ is
$f = \pi \times (0.1)^2 / 4 \pi = 2.0 \times 10^{-3}$. Assuming there are one million Sun-like
stars within 1 pc, yields 2000 stars. {\it If} the orbital eccentricities of these stars
follow the thermal distribution for $0 < e < 1$ then the fraction having $e < 0.2$
is in fact 1/25, thus giving us roughly 80 stars. {\it However} this number could be larger
for a number of reasons. The current estimates of enclosed mass at 1pc are closer
to $4 \times 10^6$ M$_\odot$ with the figure rising to over 
$ 10^7$ M$_\odot$ at less than 2 pc 
\citep{2014CQGra..31x4007S} so the number of 
stars could be closer to $10^7$. Also, as discussed above, the evolution of
orbital inclinations and/or eccentricities could well enhance the rate significantly
as stars originally on orbits inclined to the accretion disc will sink into the disc
\citep[e.g.][]{1993ApJ...409..592A}.
Also the orbits of stars located within the disc will tend to circularise over time 
\citep[e.g.][]{1993ApJ...419..166A}.

Migration of planets within protoplanetary discs 
can be extremely important. In the same way, stars may migrate within 
the accretion disc around the supermassive black hole. 
The {\it migration timescale} can be estimated as 
\citep[][]{2014MNRAS.444.2031P}:

\begin{equation}
t_{\rm mig} = {1 \over 4}{ h^2 \over q_{\rm d} q_\star } t_{\rm orb}
\end{equation}
where $h=H/R$, $q_{\rm d} = M_{\rm disc} / M_{\rm smbh}$, and $q_\star = 
M_\star/ M_{\rm smbh}$. This gives  migration timescales below 1 Myr for our
accretion disc. However, migration is a more complex issue, including the effects of
corotation torques and interactions, response of disc etc; we may find that the growing stars
 move outwards rather than inwards. One should also recall that most of the massive
stars are found within 0.4 pc, so migration might transport them inwards from around 1 pc.

One can also consider the interaction between the accreting gas and the star receiving the
material. Explicitly, one can consider whether the accretion flow ever gets close to being
{\it Eddington limited}. For the accretion histories considered here, using Equation (5), 
one can consider the mass dependence of three luminosities: the luminosity of the star
itself, $L_\star$; the luminosity released as a result of accretion of gas, $L_{\rm acc}$; 
and the Eddington luminosity, $L_{\rm Edd}$.
The point is that the accretion will be inhibited and reduced from the value expected in 
Equation (5) when $L_\star + L_{\rm acc} \ge L_{\rm Edd}$.

\begin{equation}
L_{\rm acc} = 3.14 \times 10^7 { ( \dot{M}_\star / {\rm M}_\odot {\rm yr}^{-1} ) 
\left( M_\star / {\rm M}_\odot \right) \over \left( R_\star / {\rm R}_\odot \right) }
 \ {\rm L}_\odot
\end{equation}

\begin{equation}
L_{\rm Edd} = { 4 \pi c G M_\star \over \kappa } = 3.2 \times 10^4 {\left( M_\star \over {\rm M}_\odot \right)}  {\left( \kappa_{\rm es}  \over \kappa \right) } 
 \ {\rm L}_\odot
\end{equation}

\begin{figure}
\vspace*{-0.5 cm}
\begin{center}
\includegraphics[width=2.85in]{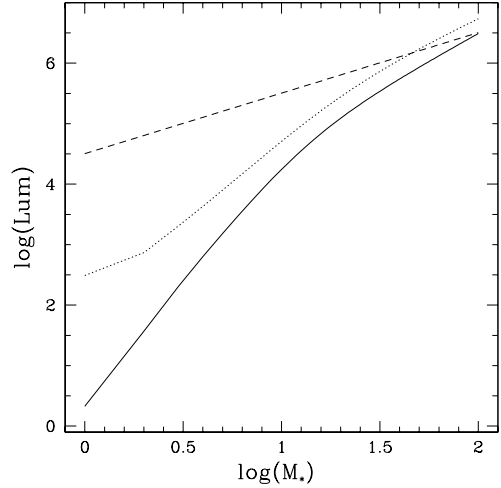} 
 \vspace*{-1.15 cm}
 \caption{Log of stellar luminosity Log($L_\star$) -- solid line, Log of Eddington 
 luminosity Log($L_{\rm Edd}$) -- dashed line, and 
 the log of the sum of stellar luminosity and accretion luminosity, Log($L_\star + L_{\rm acc}$) -- dotted
 line, plotted as a function of Log($M_\star$) for an evolution 
producing a very massive star with a very high value of $K$ (where 
$\dot{M}_\star = K M_\star^2$) 
corresponding to $t_{\rm int} =  10^5$ yr, comparable to smallest values seen in Figs. 1 and 2.}
   \label{mbdavies_figure6}
\end{center}
\end{figure}

We use the SSE fitting formulae for $L_\star$ and $R_\star$ 
\citep[][]{1996MNRAS.281..257T,2000MNRAS.315..543H}.
In Fig.\ 6 we plot $L_\star$, $L_{\rm Edd}$ and $L_\star + L_{\rm acc}$
as a function of $M_\star$ for an evolution 
producing very massive stars with a value of $K$ (where $\dot{M_\star} = K M_\star^2$) at the upper
end of what we would expect (corresponding to $t_{\rm int} = 10^5$ yr
for a star of initial mass 1 M$_\odot$).  In such a case,  
$L_\star + L_{\rm acc} \ge L_{\rm Edd}$ for $M_\star \ge
40$ M$_\odot$. Even for lower values of $K$, we would expect
the stellar mass to be limited when the stellar luminosity approaches
the Eddington luminosity once $M_\star \simeq
100$ M$_\odot$.  Hence we expect the growth of the star to be inhibited.
For the accretion disc used here, we find this to be the case for about 25 per cent
of the very massive stars produced, where the accretion rates (and thus $K$) are
sufficiently large that the accretion luminosity inhibits inflow already at stellar masses
around 40 M$_\odot$. Sufficiently massive objects could open gaps in the disc. 
Indeed gap opening may limit the growth of stellar masses in most cases, where the value
of $K$  is sufficiently low that the accretion luminosity is unlikely to inhibit the inflow
of gas onto the star.

On a related note, we comment on the likely evolution of any compact objects 
(stellar-mass black holes, white dwarfs and neutron stars). Because these objects
have much smaller radii than main-sequence stars, a given mass accretion 
rate will yield a much larger accretion luminosity (as can be seen from Equation [9]).
Equivalently, this means that accretion will be inhibited more often as the Eddington
limit will be reached at much lower mass accretion rates ($\dot{M}_\star \sim 10^{-7}$
 M$_\odot$ yr$^{-1}$) for a 10 M$_\odot$ black hole. If accretion rates are limited to this
  limit then it means that 
 {\it compact objects will not increase their mass significantly over the 10 Myr evolution
 considered here} rather the most-massive objects will form first as stars which then
 evolve to collapse forming black holes.

As we have seen above, about 100 very massive stars are likely to be
produced from low-mass stellar seeds within an accretion disc around
the supermassive black hole in the Galactic centre.  The majority of these
stars will likely evolve to produce stellar-mass black holes in supernovae. How many  
accretion discs have been produced from tidally shredded GMCs over the entire history of the 
Galactic centre? We do not know, but observations of the stellar population 
within the Galactic centre suggest that a second episode has not occurred within the
last 100 Myr \citep[][]{2011ApJ...741..108P}. But a frequency of one disc per 100 Myr 
would produce something like $10^4$ stellar-mass black holes residing within the central
parsec today. This population is roughly ten times larger than one would expect
from a regular IMF.

It should be noted that many of the massive stars in the galactic centre
are observed to be on eccentric orbits
\citep{2006ApJ...643.1011P,2006ApJ...648..405B,2009ApJ...690.1463L}.
If the stars which grew within the disc are those having low orbital eccentricities, then
one might reasonably have expected these stars to be on rather circular orbits
today. However, the system of massive stars produced around the supermassive
black hole will behave in a manner similar to an unstable planetary system. Indeed
equivalent planetary systems have been studied and shown to become unstable
on usefully short timescales \citep{1996Icar..119..261C}. In such a system, planetary
orbits evolve to the point where the orbits cross and planet-planet scattering occurs.
Scattering will leave the planets, or in this case the massive stars, on eccentric 
orbits \citep[e.g. see Fig.\,1 of][]{2020MNRAS.492..352K}.

One should note that the process we discuss in this paper is very closely-related
to what would occur within the longer-lived accretion discs found in AGN. Indeed
the production of very massive stars within AGN discs has already been discussed
\citep[e.g.][]{2003MNRAS.339..937G,2004ApJ...608..108G}.
For both AGN and less-active nuclei such as our Galactic centre, the
population of massive stars produced within the accretion discs will also
chemically enrich the central regions with ejecta from their supernova explosions
\citep{1993ApJ...409..592A}.

The stellar-mass black holes produced from the massive stars could have a number of
important roles. They may sink within the nuclear stellar cluster (kinematically) heating other stars in the
process. The enhanced population will increase the rate of so-called extreme mass-ratio inspiral
events (EMRIs) where stellar-mass black holes are captured by the supermassive black hole
producing inspiralling events potentially visible by the future LISA mission. Black holes within the 
accretion disc may also encounter each other, as they migrate within the disc, forming black hole
binaries \citep[e.g.][]{2019ApJ...878...85S}. These binaries could spiral together 
and merge and be visible with LIGO/VIRGO through their gravitational wave emission
\citep[e.g.][]{2007MNRAS.374..515L,2017MNRAS.464..946S,2017ApJ...835..165B,2018ApJ...866...66M}.
 Indeed this channel could turn out to be the favoured pathway, at least for mergers involving
the most massive (stellar mass) black holes. 
It has also been noted that evolution within the disc could ultimately produce
a more massive (intermediate-mass) black hole 
\citep[][]{2012MNRAS.425..460M,2014MNRAS.441..900M,2016ApJ...819L..17B}.

\section{Conclusions}

We have shown here that very massive stars could be produced via accretion onto low-mass stars within 
a gaseous disc surrounding the supermassive black hole in the Galactic centre. This pathway
offers an explanation for the origin of the apparently young population of  approximately 200 
massive stars found within 
$\sim 0.4$ pc of the supermassive black hole. Given the presence of molecular gas
within the central regions, it is not unreasonable to have a GMC encounter the supermassive 
black hole, and be tidally shredded, forming a disc, once every 100 Myr or so. 
Thus over time, the very central regions of the Galaxy will become enriched with stellar-mass
black holes. The supernovae that produced them will also chemically enrich the region.
Black holes may form binaries which merge and are potential gravitational-wave sources.
The merger rate between stellar-mass black holes and the supermassive black hole 
will also be enhanced.  As has been pointed out by others, similar processes as we describe for
the Galactic centre will also likely occur within the accretion discs of AGN.













\section*{Acknowledgements}

We acknowledge the referee for their  extremely helpful referee report.
MBD acknowledges useful discussions with Ross Church and Anders Johansen. MBD is supported
by the project grant 2014.0017 ``IMPACT" from the Knut and Alice Wallenberg Foundation.


\section*{Data Availability}

The data underlying this article will be shared on reasonable request to the corresponding author.

\bibliographystyle{mnras}
\bibliography{mms}

\begin{thebibliography}{}
\makeatletter
\relax
\def\mn@urlcharsother{\let\do\@makeother \do\$\do\&\do\#\do\^\do\_\do\%\do\~}
\def\mn@doi{\begingroup\mn@urlcharsother \@ifnextchar [ {\mn@doi@}
  {\mn@doi@[]}}
\def\mn@doi@[#1]#2{\def\@tempa{#1}\ifx\@tempa\@empty \href
  {http://dx.doi.org/#2} {doi:#2}\else \href {http://dx.doi.org/#2} {#1}\fi
  \endgroup}
\def\mn@eprint#1#2{\mn@eprint@#1:#2::\@nil}
\def\mn@eprint@arXiv#1{\href {http://arxiv.org/abs/#1} {{\tt arXiv:#1}}}
\def\mn@eprint@dblp#1{\href {http://dblp.uni-trier.de/rec/bibtex/#1.xml}
  {dblp:#1}}
\def\mn@eprint@#1:#2:#3:#4\@nil{\def\@tempa {#1}\def\@tempb {#2}\def\@tempc
  {#3}\ifx \@tempc \@empty \let \@tempc \@tempb \let \@tempb \@tempa \fi \ifx
  \@tempb \@empty \def\@tempb {arXiv}\fi \@ifundefined
  {mn@eprint@\@tempb}{\@tempb:\@tempc}{\expandafter \expandafter \csname
  mn@eprint@\@tempb\endcsname \expandafter{\@tempc}}}

\bibitem[\protect\citeauthoryear{{Alexander}, {Begelman}  \&
  {Armitage}}{{Alexander} et~al.}{2007}]{2007ApJ...654..907A}
{Alexander} R.~D.,  {Begelman} M.~C.,   {Armitage} P.~J.,  2007, \mn@doi [\apj]
  {10.1086/509709}, \href
  {https://ui.adsabs.harvard.edu/abs/2007ApJ...654..907A} {654, 907}

\bibitem[\protect\citeauthoryear{{Artymowicz}}{{Artymowicz}}{1993}]{1993ApJ...419..166A}
{Artymowicz} P.,  1993, \mn@doi [\apj] {10.1086/173470}, \href
  {https://ui.adsabs.harvard.edu/abs/1993ApJ...419..166A} {419, 166}

\bibitem[\protect\citeauthoryear{{Artymowicz}, {Lin}  \&
  {Wampler}}{{Artymowicz} et~al.}{1993}]{1993ApJ...409..592A}
{Artymowicz} P.,  {Lin} D.~N.~C.,   {Wampler} E.~J.,  1993, \mn@doi [\apj]
  {10.1086/172690}, \href
  {https://ui.adsabs.harvard.edu/abs/1993ApJ...409..592A} {409, 592}

\bibitem[\protect\citeauthoryear{{Bartko} et~al.,}{{Bartko}
  et~al.}{2010}]{2010ApJ...708..834B}
{Bartko} H.,  et~al., 2010, \mn@doi [\apj] {10.1088/0004-637X/708/1/834}, \href
  {https://ui.adsabs.harvard.edu/abs/2010ApJ...708..834B} {708, 834}

\bibitem[\protect\citeauthoryear{{Bartos}, {Kocsis}, {Haiman}  \&
  {M{\'a}rka}}{{Bartos} et~al.}{2017}]{2017ApJ...835..165B}
{Bartos} I.,  {Kocsis} B.,  {Haiman} Z.,   {M{\'a}rka} S.,  2017, \mn@doi
  [\apj] {10.3847/1538-4357/835/2/165}, \href
  {https://ui.adsabs.harvard.edu/abs/2017ApJ...835..165B} {835, 165}

\bibitem[\protect\citeauthoryear{{Bellovary}, {Mac Low}, {McKernan}  \&
  {Ford}}{{Bellovary} et~al.}{2016}]{2016ApJ...819L..17B}
{Bellovary} J.~M.,  {Mac Low} M.-M.,  {McKernan} B.,   {Ford} K.~E.~S.,  2016,
  \mn@doi [\apjl] {10.3847/2041-8205/819/2/L17}, \href
  {https://ui.adsabs.harvard.edu/abs/2016ApJ...819L..17B} {819, L17}

\bibitem[\protect\citeauthoryear{{Beloborodov}, {Levin}, {Eisenhauer},
  {Genzel}, {Paumard}, {Gillessen}  \& {Ott}}{{Beloborodov}
  et~al.}{2006}]{2006ApJ...648..405B}
{Beloborodov} A.~M.,  {Levin} Y.,  {Eisenhauer} F.,  {Genzel} R.,  {Paumard}
  T.,  {Gillessen} S.,   {Ott} T.,  2006, \mn@doi [\apj] {10.1086/504279},
  \href {https://ui.adsabs.harvard.edu/abs/2006ApJ...648..405B} {648, 405}

\bibitem[\protect\citeauthoryear{{Bonnell} \& {Rice}}{{Bonnell} \&
  {Rice}}{2008}]{2008Sci...321.1060B}
{Bonnell} I.~A.,  {Rice} W.~K.~M.,  2008, \mn@doi [Science]
  {10.1126/science.1160653}, \href
  {https://ui.adsabs.harvard.edu/abs/2008Sci...321.1060B} {321, 1060}

\bibitem[\protect\citeauthoryear{{Chambers}, {Wetherill}  \& {Boss}}{{Chambers}
  et~al.}{1996}]{1996Icar..119..261C}
{Chambers} J.~E.,  {Wetherill} G.~W.,   {Boss} A.~P.,  1996, \mn@doi [\icarus]
  {10.1006/icar.1996.0019}, \href
  {https://ui.adsabs.harvard.edu/abs/1996Icar..119..261C} {119, 261}

\bibitem[\protect\citeauthoryear{{Edgar}}{{Edgar}}{2004}]{2004NewAR..48..843E}
{Edgar} R.,  2004, \mn@doi [\nar] {10.1016/j.newar.2004.06.001}, \href
  {https://ui.adsabs.harvard.edu/abs/2004NewAR..48..843E} {48, 843}

\bibitem[\protect\citeauthoryear{{Gammie}}{{Gammie}}{2001}]{2001ApJ...553..174G}
{Gammie} C.~F.,  2001, \mn@doi [\apj] {10.1086/320631}, \href
  {https://ui.adsabs.harvard.edu/abs/2001ApJ...553..174G} {553, 174}

\bibitem[\protect\citeauthoryear{{Genzel}, {Pichon}, {Eckart}, {Gerhard}  \&
  {Ott}}{{Genzel} et~al.}{2000}]{2000MNRAS.317..348G}
{Genzel} R.,  {Pichon} C.,  {Eckart} A.,  {Gerhard} O.~E.,   {Ott} T.,  2000,
  \mn@doi [\mnras] {10.1046/j.1365-8711.2000.03582.x}, \href
  {https://ui.adsabs.harvard.edu/abs/2000MNRAS.317..348G} {317, 348}

\bibitem[\protect\citeauthoryear{{Genzel}, {Eisenhauer}  \&
  {Gillessen}}{{Genzel} et~al.}{2010}]{2010RvMP...82.3121G}
{Genzel} R.,  {Eisenhauer} F.,   {Gillessen} S.,  2010, \mn@doi [Reviews of
  Modern Physics] {10.1103/RevModPhys.82.3121}, \href
  {https://ui.adsabs.harvard.edu/abs/2010RvMP...82.3121G} {82, 3121}

\bibitem[\protect\citeauthoryear{{Gerhard}}{{Gerhard}}{2001}]{2001ApJ...546L..39G}
{Gerhard} O.,  2001, \mn@doi [\apjl] {10.1086/318054}, \href
  {https://ui.adsabs.harvard.edu/abs/2001ApJ...546L..39G} {546, L39}

\bibitem[\protect\citeauthoryear{{Goodman}}{{Goodman}}{2003}]{2003MNRAS.339..937G}
{Goodman} J.,  2003, \mn@doi [\mnras] {10.1046/j.1365-8711.2003.06241.x}, \href
  {https://ui.adsabs.harvard.edu/abs/2003MNRAS.339..937G} {339, 937}

\bibitem[\protect\citeauthoryear{{Goodman} \& {Tan}}{{Goodman} \&
  {Tan}}{2004}]{2004ApJ...608..108G}
{Goodman} J.,  {Tan} J.~C.,  2004, \mn@doi [\apj] {10.1086/386360}, \href
  {https://ui.adsabs.harvard.edu/abs/2004ApJ...608..108G} {608, 108}

\bibitem[\protect\citeauthoryear{{Hobbs} \& {Nayakshin}}{{Hobbs} \&
  {Nayakshin}}{2009}]{2009MNRAS.394..191H}
{Hobbs} A.,  {Nayakshin} S.,  2009, \mn@doi [\mnras]
  {10.1111/j.1365-2966.2008.14359.x}, \href
  {https://ui.adsabs.harvard.edu/abs/2009MNRAS.394..191H} {394, 191}

\bibitem[\protect\citeauthoryear{{Hurley}, {Pols}  \& {Tout}}{{Hurley}
  et~al.}{2000}]{2000MNRAS.315..543H}
{Hurley} J.~R.,  {Pols} O.~R.,   {Tout} C.~A.,  2000, \mn@doi [\mnras]
  {10.1046/j.1365-8711.2000.03426.x}, \href
  {https://ui.adsabs.harvard.edu/abs/2000MNRAS.315..543H} {315, 543}

\bibitem[\protect\citeauthoryear{{Kokaia}, {Davies}  \& {Mustill}}{{Kokaia}
  et~al.}{2020}]{2020MNRAS.492..352K}
{Kokaia} G.,  {Davies} M.~B.,   {Mustill} A.~J.,  2020, \mn@doi [\mnras]
  {10.1093/mnras/stz3408}, \href
  {https://ui.adsabs.harvard.edu/abs/2020MNRAS.492..352K} {492, 352}

\bibitem[\protect\citeauthoryear{{Levin}}{{Levin}}{2007}]{2007MNRAS.374..515L}
{Levin} Y.,  2007, \mn@doi [\mnras] {10.1111/j.1365-2966.2006.11155.x}, \href
  {https://ui.adsabs.harvard.edu/abs/2007MNRAS.374..515L} {374, 515}

\bibitem[\protect\citeauthoryear{{Levin} \& {Beloborodov}}{{Levin} \&
  {Beloborodov}}{2003}]{2003ApJ...590L..33L}
{Levin} Y.,  {Beloborodov} A.~M.,  2003, \mn@doi [\apjl] {10.1086/376675},
  \href {https://ui.adsabs.harvard.edu/abs/2003ApJ...590L..33L} {590, L33}

\bibitem[\protect\citeauthoryear{{Lu}, {Ghez}, {Hornstein}, {Morris}, {Becklin}
   \& {Matthews}}{{Lu} et~al.}{2009}]{2009ApJ...690.1463L}
{Lu} J.~R.,  {Ghez} A.~M.,  {Hornstein} S.~D.,  {Morris} M.~R.,  {Becklin}
  E.~E.,   {Matthews} K.,  2009, \mn@doi [\apj] {10.1088/0004-637X/690/2/1463},
  \href {https://ui.adsabs.harvard.edu/abs/2009ApJ...690.1463L} {690, 1463}

\bibitem[\protect\citeauthoryear{{Lucas}, {Bonnell}, {Davies}  \&
  {Rice}}{{Lucas} et~al.}{2013}]{2013MNRAS.433..353L}
{Lucas} W.~E.,  {Bonnell} I.~A.,  {Davies} M.~B.,   {Rice} W.~K.~M.,  2013,
  \mn@doi [\mnras] {10.1093/mnras/stt727}, \href
  {https://ui.adsabs.harvard.edu/abs/2013MNRAS.433..353L} {433, 353}

\bibitem[\protect\citeauthoryear{{MacLeod} \& {Lin}}{{MacLeod} \&
  {Lin}}{2020}]{2020ApJ...889...94M}
{MacLeod} M.,  {Lin} D. N.~C.,  2020, \mn@doi [\apj]
  {10.3847/1538-4357/ab64db}, \href
  {https://ui.adsabs.harvard.edu/abs/2020ApJ...889...94M} {889, 94}

\bibitem[\protect\citeauthoryear{{McKernan}, {Ford}, {Lyra}  \&
  {Perets}}{{McKernan} et~al.}{2012}]{2012MNRAS.425..460M}
{McKernan} B.,  {Ford} K.~E.~S.,  {Lyra} W.,   {Perets} H.~B.,  2012, \mn@doi
  [\mnras] {10.1111/j.1365-2966.2012.21486.x}, \href
  {https://ui.adsabs.harvard.edu/abs/2012MNRAS.425..460M} {425, 460}

\bibitem[\protect\citeauthoryear{{McKernan}, {Ford}, {Kocsis}, {Lyra}  \&
  {Winter}}{{McKernan} et~al.}{2014}]{2014MNRAS.441..900M}
{McKernan} B.,  {Ford} K.~E.~S.,  {Kocsis} B.,  {Lyra} W.,   {Winter} L.~M.,
  2014, \mn@doi [\mnras] {10.1093/mnras/stu553}, \href
  {https://ui.adsabs.harvard.edu/abs/2014MNRAS.441..900M} {441, 900}

\bibitem[\protect\citeauthoryear{{McKernan} et~al.,}{{McKernan}
  et~al.}{2018}]{2018ApJ...866...66M}
{McKernan} B.,  et~al., 2018, \mn@doi [\apj] {10.3847/1538-4357/aadae5}, \href
  {https://ui.adsabs.harvard.edu/abs/2018ApJ...866...66M} {866, 66}

\bibitem[\protect\citeauthoryear{{Nayakshin}}{{Nayakshin}}{2006}]{2006MNRAS.372..143N}
{Nayakshin} S.,  2006, \mn@doi [\mnras] {10.1111/j.1365-2966.2006.10772.x},
  \href {https://ui.adsabs.harvard.edu/abs/2006MNRAS.372..143N} {372, 143}

\bibitem[\protect\citeauthoryear{{Nayakshin} \& {Cuadra}}{{Nayakshin} \&
  {Cuadra}}{2005}]{2005A&A...437..437N}
{Nayakshin} S.,  {Cuadra} J.,  2005, \mn@doi [\aap]
  {10.1051/0004-6361:20042052}, \href
  {https://ui.adsabs.harvard.edu/abs/2005A&A...437..437N} {437, 437}

\bibitem[\protect\citeauthoryear{{Nayakshin}, {Cuadra}  \&
  {Springel}}{{Nayakshin} et~al.}{2007}]{2007MNRAS.379...21N}
{Nayakshin} S.,  {Cuadra} J.,   {Springel} V.,  2007, \mn@doi [\mnras]
  {10.1111/j.1365-2966.2007.11938.x}, \href
  {https://ui.adsabs.harvard.edu/abs/2007MNRAS.379...21N} {379, 21}

\bibitem[\protect\citeauthoryear{{Paardekooper}}{{Paardekooper}}{2014}]{2014MNRAS.444.2031P}
{Paardekooper} S.~J.,  2014, \mn@doi [\mnras] {10.1093/mnras/stu1542}, \href
  {https://ui.adsabs.harvard.edu/abs/2014MNRAS.444.2031P} {444, 2031}

\bibitem[\protect\citeauthoryear{{Papaloizou} \& {Lin}}{{Papaloizou} \&
  {Lin}}{1995}]{1995ARA&A..33..505P}
{Papaloizou} J.~C.~B.,  {Lin} D.~N.~C.,  1995, \mn@doi [\araa]
  {10.1146/annurev.aa.33.090195.002445}, \href
  {https://ui.adsabs.harvard.edu/abs/1995ARA&A..33..505P} {33, 505}

\bibitem[\protect\citeauthoryear{{Paumard} et~al.,}{{Paumard}
  et~al.}{2006}]{2006ApJ...643.1011P}
{Paumard} T.,  et~al., 2006, \mn@doi [\apj] {10.1086/503273}, \href
  {https://ui.adsabs.harvard.edu/abs/2006ApJ...643.1011P} {643, 1011}

\bibitem[\protect\citeauthoryear{{Pfuhl} et~al.,}{{Pfuhl}
  et~al.}{2011}]{2011ApJ...741..108P}
{Pfuhl} O.,  et~al., 2011, \mn@doi [\apj] {10.1088/0004-637X/741/2/108}, \href
  {https://ui.adsabs.harvard.edu/abs/2011ApJ...741..108P} {741, 108}

\bibitem[\protect\citeauthoryear{{Sch{\"o}del}, {Feldmeier}, {Neumayer},
  {Meyer}  \& {Yelda}}{{Sch{\"o}del} et~al.}{2014}]{2014CQGra..31x4007S}
{Sch{\"o}del} R.,  {Feldmeier} A.,  {Neumayer} N.,  {Meyer} L.,   {Yelda} S.,
  2014, \mn@doi [Classical and Quantum Gravity]
  {10.1088/0264-9381/31/24/244007}, \href
  {https://ui.adsabs.harvard.edu/abs/2014CQGra..31x4007S} {31, 244007}

\bibitem[\protect\citeauthoryear{{Secunda}, {Bellovary}, {Mac Low}, {Ford},
  {McKernan}, {Leigh}, {Lyra}  \& {S{\'a}ndor}}{{Secunda}
  et~al.}{2019}]{2019ApJ...878...85S}
{Secunda} A.,  {Bellovary} J.,  {Mac Low} M.-M.,  {Ford} K.~E.~S.,  {McKernan}
  B.,  {Leigh} N. W.~C.,  {Lyra} W.,   {S{\'a}ndor} Z.,  2019, \mn@doi [\apj]
  {10.3847/1538-4357/ab20ca}, \href
  {https://ui.adsabs.harvard.edu/abs/2019ApJ...878...85S} {878, 85}

\bibitem[\protect\citeauthoryear{{Stone}, {Metzger}  \& {Haiman}}{{Stone}
  et~al.}{2017}]{2017MNRAS.464..946S}
{Stone} N.~C.,  {Metzger} B.~D.,   {Haiman} Z.,  2017, \mn@doi [\mnras]
  {10.1093/mnras/stw2260}, \href
  {https://ui.adsabs.harvard.edu/abs/2017MNRAS.464..946S} {464, 946}

\bibitem[\protect\citeauthoryear{{Su}, {Slatyer}  \& {Finkbeiner}}{{Su}
  et~al.}{2010}]{2010ApJ...724.1044S}
{Su} M.,  {Slatyer} T.~R.,   {Finkbeiner} D.~P.,  2010, \mn@doi [\apj]
  {10.1088/0004-637X/724/2/1044}, \href
  {https://ui.adsabs.harvard.edu/abs/2010ApJ...724.1044S} {724, 1044}

\bibitem[\protect\citeauthoryear{{Tout}, {Pols}, {Eggleton}  \& {Han}}{{Tout}
  et~al.}{1996}]{1996MNRAS.281..257T}
{Tout} C.~A.,  {Pols} O.~R.,  {Eggleton} P.~P.,   {Han} Z.,  1996, \mn@doi
  [\mnras] {10.1093/mnras/281.1.257}, \href
  {https://ui.adsabs.harvard.edu/abs/1996MNRAS.281..257T} {281, 257}

\bibitem[\protect\citeauthoryear{{Zubovas}, {King}  \& {Nayakshin}}{{Zubovas}
  et~al.}{2011}]{2011MNRAS.415L..21Z}
{Zubovas} K.,  {King} A.~R.,   {Nayakshin} S.,  2011, \mn@doi [\mnras]
  {10.1111/j.1745-3933.2011.01070.x}, \href
  {https://ui.adsabs.harvard.edu/abs/2011MNRAS.415L..21Z} {415, L21}

\makeatother
\end{thebibliography}

\bsp    
\label{lastpage}

\end{document}